\documentclass[reprint,aps,prl,amssymb, amsmath,superscriptaddress]{revtex4-1}
\usepackage{graphicx}
\usepackage{pifont}
\usepackage{hyperref}
\hypersetup{
    colorlinks=true, 
    linktoc=all,     
    linkcolor=blue,  
    citecolor =blue
}
\usepackage[T1]{fontenc}
\usepackage[normalem]{ulem}
\let\vaccent=\v 
\renewcommand{\v}[1]{\ensuremath{\mathbf{#1}}} 
\newcommand{\avg}[1]{\langle#1\rangle}
\newcommand{\abs}[1]{\left| #1 \right|} 

\newcommand{\dbar}{d\hspace*{-0.08em}\bar{}\hspace*{0.1em}}

\begin{document}
\title{Hidden Charge  Order of Interacting Dirac Fermions 
on the Honeycomb Lattice}
\author{Elliot Christou}
\affiliation{London Centre for Nanotechnology, University College London, Gordon St., London, WC1H 0AH, United Kingdom}
\author{Bruno Uchoa}
\affiliation{Department of Physics and Astronomy, University of Oklahoma, Norman, OK 73069, USA}
\author{Frank Kr\"uger}
\affiliation{London Centre for Nanotechnology, University College London, Gordon St., London, WC1H 0AH, United Kingdom}
\affiliation{ISIS Facility, Rutherford Appleton Laboratory, Chilton, Didcot, Oxfordshire OX11 0QX, United Kingdom}

\date{\today}
  
\begin{abstract}
We consider the extended half-filled Hubbard model on the honeycomb lattice for second nearest neighbors interactions. Using a functional integral approach, we find that collective fluctuations suppress topological states and instead favor charge ordering, in agreement with previous numerical studies. However, we show that the critical point is not of the putative semimetal-Mott insulator variety. Due to the frustrated nature of the interactions, the ground state is described by a  novel hidden {\it metallic} charge order with {\it semi-Dirac} excitations. We conjecture that this transition is not in the Gross-Neveu universality class.
\end{abstract}


\maketitle
The extended, half-filled Hubbard model on the honeycomb lattice exhibits a rich phase diagram, even at  mean-field level.
 The low energy excitations in the semimetallic phase are massless Dirac fermions \cite{novoselovetalnature2005}, which
 couple  to the order-parameter fluctuations and 
 are known to change 
 the universal critical behaviour to that of the Gross-Neveu-Yukawa (GNY) \cite{grossneveuprd1974} variety.  
 For the transition from the Dirac semimetal to the antiferromagnetic Mott insulator, driven by 
 the on-site Hubbard repulsion $U$, this has been well understood  through a combination of analytical low-energy 
theories \cite{herbutprl2006, herbutetalprb2009} and sign-free auxiliary-field quantum Monte Carlo 
\cite{assaadherbutprx2013,otsukaetalprx2016,satoetalprl2017}.
 
\begin{figure}[t] 
\includegraphics[width=0.85\columnwidth]{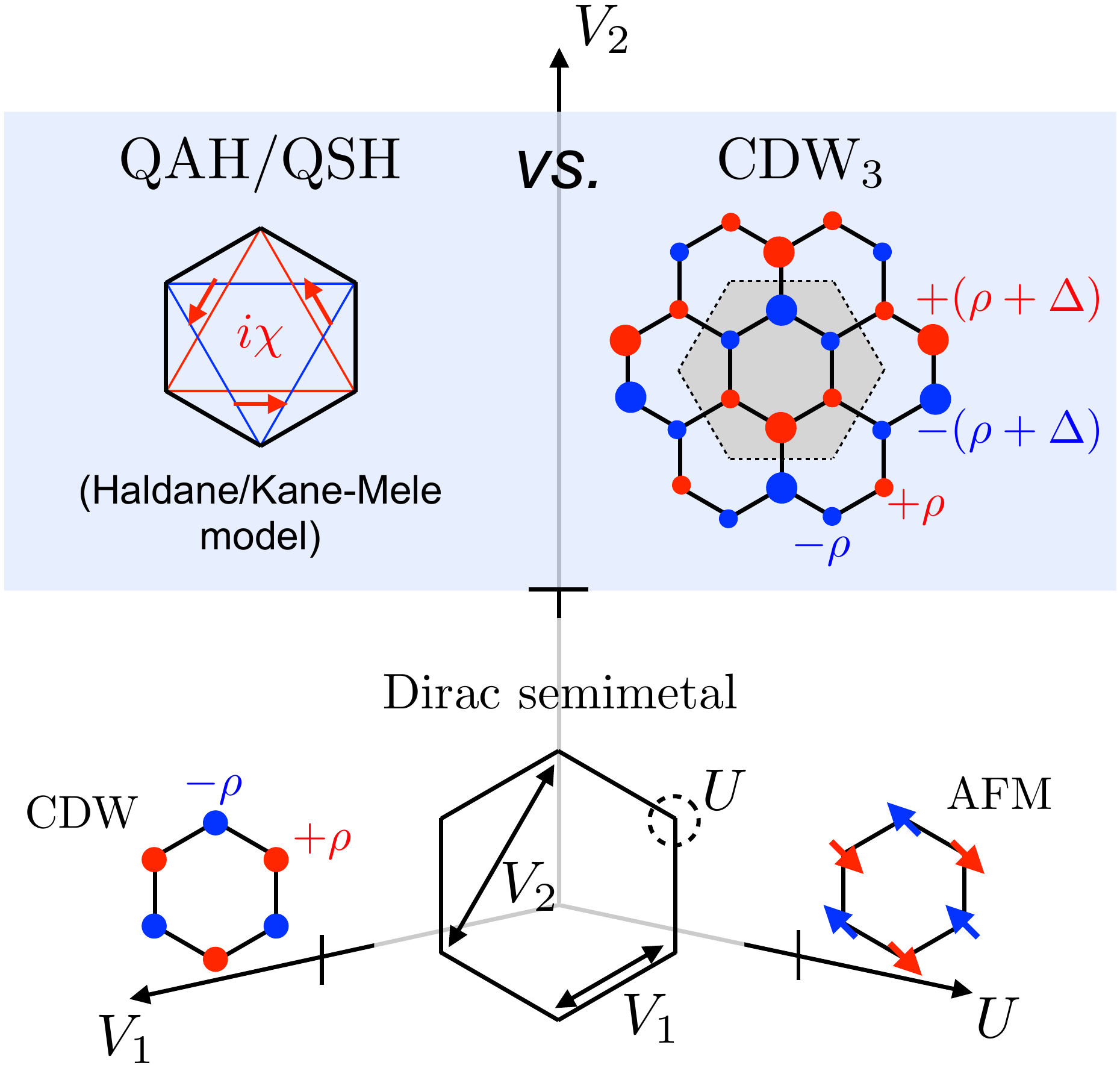}
\caption{\label{competing} (color online) Schematic phase diagram of the half-filled Hubbard model on the honeycomb lattice. 
The on-site and NN repulsions $U$ and $V_1$ induce antiferromagnetic (AFM) and 
charge-density wave (CDW) states, respectively.  At large NNN interactions $V_2$, there is phase competition between a topological 
Mott insulator and charge-ordered states with enlarged unit cell (CDW$_3$). The charge modulation is shown relative to half filling.}
\end{figure}

Of the many broken-symmetry phases driven by nearest neighbor (NN) and next-nearest neighbor (NNN) repulsions, topological phases 
are favored by strong NNN interactions ($V_2$)  \cite{raghuetalprl2008}, which can stabilize the Kane-Mele quantum spin Hall phase (QSH) 
in the spinful model \cite{kanemeleprlA2005}, or the Haldane quantum anomalous Hall (QAH) state in the spinless case \cite{haldaneprl1988}.
Those states nevertheless compete with unconventional charge order (see Fig.~\ref{competing}) that extends beyond the honeycomb unit 
cell \cite{grushinetalprb2013}. One would expect quantum fluctuations  to play a crucial role in determining the fate of the topological phases, 
in particular the soft fluctuations associated with breaking of continuous spin rotational symmetry in the QSH phase. Unfortunately, the sign 
problem for large $V_2$ prevents the use of quantum Monte Carlo methods \cite{golorwesselprb2015}. Extensive 
numerical research into spinless \cite{garciamartinezetalprb2013,daghoferhohenadlerprb2014,djuricetalprb2014,capponilauchliprb2015,motruketalprb2015, 
schereretalprb2015} and spinful \cite{volpezetalprb2016,delapenaetalprb2017, kuritaetalprb2016,bijelicetalprb2018} models using 
exact diagonalization, variational Monte Carlo, infinite density matrix RG,  and functional RG have been pivotal to determine the phase behavior.  

In this Rapid Communication, we analytically examine the role of fluctuations for the phase competition along the $V_2$ axis. We derive an effective 
low-energy description for the quantum phase transition into the charge ordered CDW$_3$ state, 
and analyze the leading instabilities in the presence of the long wavelength collective fluctuation fields via a functional integral approach. 
Our analytical results  are convincingly consistent with numerical lattice calculations, which suggest that CDW$_3$ order is favored 
over topological Mott insulating phases. However, surprisingly, we find that the onset of CDW$_3$ order does not produce a many-body Mott gap, 
but rather  a novel  hidden metallic order
as a result of the frustration of the $V_2$ interaction on the triangular sublattices. 
The low energy excitations of this state 
are massless {\it semi-Dirac} quasiparticles \cite{banerjeeetalprl2009}, which disperse linearly in one direction and parabolically in the other. 
We show that this state is robust against fluctuation effects. 
We conjecture that the phase transition to the metallic CDW$_3$ state  is not in the GNY universality class.

{\it Model and low energy description.}$-$
Our starting model is given by the Hamiltonian  
\begin{align}
\mathcal{H}&=-t\sum_{\avg{i,j}}\sum_{s=\uparrow,\downarrow}(c^\dagger_{is}c_{js}+\text{h.c.})+V_2\sum_{\avg{\avg{i,j}}}\hat{n}_i \hat{n}_j
\label{Ham1}
\end{align}
on the half-filled honeycomb lattice with NN hopping $t$ and NNN repulsion $V_2$, where $c_{is}$ is an annihilation operator for an electron with 
spin $s$ on site $i$ and  $\hat{n}_i=\hat{n}_{i\uparrow}+\hat{n}_{i\downarrow}$ the density operator. The corresponding spinless model is obtained 
by suppressing the spin index $s$. In the absence of interactions, the  low-energy theory of the semimetallic state describes massless Dirac fermions 
at the corners of the Brillouin zone $K_{\nu=\pm}=\frac{4\pi}{3\sqrt{3}}(\nu,0)$,
\begin{align} \mathcal{H}_t&=v_F\underset{{\abs{\v{p}}\le{\Lambda}}}{\int}\!\dbar^2\v{p}\,\Psi_{\v{p}}^{\dagger}s^0(p_x\sigma^x\nu^z +p_y\sigma^y\nu^0)\Psi_{\v{p}},\label{dirac}
\end{align}
where $v_F$ is the Fermi velocity, $s^\mu,\, \sigma^\mu,\, \nu^\mu$ ($\mu =0, x,y,z$) are the 4-vectors of identity and Pauli matrices acting respectively on 
the spin, sublattice, and valley spaces and 
\begin{equation}
\Psi_{\v{p}}=(\psi_{\v{p}A}^{+\uparrow}, \psi_{\v{p}A}^{-\uparrow},\psi_{\v{p}B}^{+\uparrow}, \psi_{\v{p}B}^{-\uparrow}, \psi_{\v{p}A}^{+\downarrow},\psi_{\v{p}A}^{-\downarrow}, \psi_{\v{p}B}^{+\downarrow}, \psi_{\v{p}B}^{-\downarrow})
\end{equation}
is an eight component spinor. 
The measure $\dbar^2\v{p}=d^2\v{p}N/A$, with $\,A=2\pi\Lambda^2$, conserves the number of states $N$ between the lattice and effective models,  where  $\Lambda$  is the ultraviolet cut-off.

Decomposition of Hamiltonian (\ref{Ham1}) in the bond-order order channel,  $\hat{\chi}^{\mu}_{ij}=c^\dagger_is^{\mu}c_j$, gives the effective description of the topological Mott insulator states  \cite{raghuetalprl2008}. 
 Enacting the mean-field decoupling in this channel and imposing a 
 translationally invariant, sublattice dependent, and purely imaginary \textit{ansatz} $\avg{\hat{\chi}^\mu_{ij}}=i\chi^\mu\sigma^z$, which is known to 
 minimize the free energy \cite{raghuetalprl2008}, the effective mass terms are 
\begin{align}
\mathcal{H}_\chi=3V_2\left(\chi^{\mu}\chi^{\mu}+\frac{\sqrt{3}}{2}
\int\dbar^2\v{p}\,
\Psi_{\v{p}}^{\dagger}\chi^\mu s^{\mu}\sigma^z\nu^z\Psi_{\v{p}}\right),\label{topologicalmass}
\end{align}   
where summation of repeated $\mu$ indices is implied. The singlet ($\mu=0$) component of $\chi^\mu$  describes the order parameter of the QAH phase, 
which spontaneously breaks global time reversal symmetry, opening a Mott gap at the Dirac points. Similarly, a non-zero triplet component ($\mu\neq 0$) 
describes the QSH state, which 
 spontaneously breaks $SU(2)$ spin-rotational symmetry  but preserves time reversal symmetry. The electron mean field dispersion takes the same form 
 in the QAH and QSH phases, $|\varepsilon_{s\nu} (\v{p})|=\sqrt{v_F^2\abs{\v{p}}^2 + \left(3\sqrt{3}V_2 \chi^\mu/2 \right)^2}.\label{disp1}$

To describe the competing CDW$_3$ phase (Fig.~\ref{competing}), we decouple the interaction in the 
density channel and apply the plaquette \textit{ansatz} \cite{grushinetalprb2013} for the charge occupation $\avg{\hat{n}_i}=\rho_0 + \rho_i$ which describes 
the deviation of charge occupation $\{\rho_i\}=\{\rho,-\rho,-(\rho+\Delta),-\rho,\rho,\rho+\Delta\}$ from the half filling value $\rho_0=N_s/2$
(where $N_s=1\text{ or }2$ is the number of fermionic spin flavors).
 In total, there are 9 equivalent configurations of the CDW$_3$ state related by $2\pi/3$ rotations and translations \cite{garciamartinezetalprb2013}.
The constraints $0\le\Delta\le\rho\le\rho_0$ and $\rho+\Delta\le\rho_0$ ensure the filling is devoid of pathology.
Such a phase spontaneously breaks translational symmetry and keeps only one mirror: $C_{6v}\rightarrow{}C_{1v}$.

The CDW$_3$ phase is characterized by an enlarged 6-site unit cell covering an entire honeycomb plaquette (Fig.~\ref{competing}). 
The resultant down-folding of the bands increases the number of energy levels at a given momentum threefold. That gives rise to six bands 
with an additional 2-fold degeneracy in the spinful model, and maps the Dirac points onto the $\Gamma$ point ($\v{p}=\v{0}$), as shown in 
Fig.~\ref{semi-Dirac fermions+folded BZ}(a). Integrating out the high
energy bands (see Supplemental Material \cite{supplementalmaterial}), the interaction part of the Hamiltonian 
$\tilde{\mathcal{H}} = \tilde{\mathcal{H}}_t+\tilde{\mathcal{H}}_\delta$ in the projected space reads 

\begin{align}
  \tilde{\mathcal{H}}_\delta &= \int \! \tilde{\Psi}_{\v{p}}^{\dagger} s^0\Big\{\delta_1\tau^z \tau^0+\frac{\delta_2}{2}[C_n \tau^0( S_m \tau^x -C_m\tau^y)\notag\\
&\phantom{=}\, -S_n \tau^z(C_m\tau^x+S_m\tau^y)]\Big\}\tilde{\Psi}_{\v{p}} + E_0,
\end{align}
with $\delta_1 = 2 V_2(\rho-\Delta),\,\delta_2=2V_2(\rho+\Delta/2)$, $E_0 = (4\delta_2^2-\delta_1^2)/6V_2$ . Here, $C_n=\cos( 2\pi n /3)$, $S_n=\sin( 2\pi n /3)$ and $n,m=1,2,3$ enumerate the 9 possible broken symmetry state configurations. Written as a combination of irreducible representations \cite{Basko08, dejuanprb2013}, order parameter $\delta_1$ couples to the charge imbalance between the $A$ and $B$ sublattices, whereas $\delta_2$ couples to the broken rotations ($n$)  and translations ($m$) of each configuration. The energy dispersion is degenerate up to a $2\pi/3$ rotation, and hence all configurations have the same free energy. In the following, we refer to the $(n,m)=(3,1)$ pattern in Fig.~\ref{competing}. 

In the projected space, $\tilde{\mathcal{H}}_t$ 
has the same form as in (\ref{dirac}) adopting the substitution  $\vec{\sigma}\otimes\vec{\nu}\to \vec{\tau}\otimes\vec{\tau}$ to represent the effective, four-dimensional 
low-energy theory after down-folding and projection. The resulting mean-field dispersion is given by
\begin{equation}
|\tilde{\varepsilon}_{s,\pm}(\v{p})|= \sqrt{v^2_F\abs{\v{p}}^2+\delta_1^2+\delta_2^2\pm2\delta_2\sqrt{v^2_Fp^2_y+\delta_1^2}}.\label{disp2}
\end{equation}

{\it Mean-field phase diagram.}$-$ We expand the Ginzburg Landau free energy density in terms of the different order parameters. Since there is no evidence for phase 
coexistence we analyze the cases of QAH/QSH order and CDW$_3$ order separately. This is sufficient to identify the leading instability 
along the $V_2$ axis.  For the topological Mott insulators we obtain the free-energy expansion 
\begin{equation}
f_{\text{mf}}(\chi)=\alpha_{\text{mf}} \chi^2 + \beta_{\text{mf}} \abs{\chi}^3
\end{equation}
 with $\chi=\chi^0$ and $\chi=\chi^z$ in the QAH and QSH phases, respectively.  The mean-field coefficients do not depend on the channel in which 
the symmetry is broken, indicating that at this level, the QAH and QSH phases are degenerate. Note that the presence of a stabilizing cubic term 
in the free energy is generic for Dirac fermions \cite{Sachdev}. For the quadratic coefficient we obtain $\alpha_{\text{mf}}=3V_2(1-9v_2)$ with 
$v_2 = \pi\Lambda V_2/v_FA$, indicating a continuous phase transition between the semimetal and a topological Mott insulator at a critical coupling 
$(v_2)_c=1/9$.

The analysis is more involved for the CDW$_3$ state due to the two-gap structure $\delta_1$ and $\delta_2$. Using the parametrization $\Delta=x\rho$ for $0\le{}x\le1$ 
we obtain 
\begin{equation}
f_\textrm{mf}(x,\rho)= \tilde{\alpha}_\mathrm{mf}(x) \rho^2 +\tilde{\beta}_\mathrm{mf}(x) \abs{\rho}^3,
\end{equation} 
where $\tilde{\alpha}_{\text{mf}}(x)=2 V_2[1+2x-6N_sv_2(1-x+\frac{3}{4}x^2)]$ and $\tilde{\beta}_{\text{mf}}(x)=8\pi{}N_sv_2^2 V_2 (2-3x+x^3)$, with $N_s= 1, 2$ the 
spin degeneracy \cite{note4}.  By inspection, the CDW$_3$ state with $x=0$ ($\Delta=0$) is the leading instability at a critical coupling $(\tilde{v}_2)_c = 1/(6N_s)$. 
In the ordered phase, the $\Delta=0$ state remains energetically favorable until large values of $V_2$ outside the range of applicability of the model. 

To summarize,  for the spinless case ($N_s=1$), the topological QAH Mott insulator is the leading instability at a critical 
coupling $(v_2)_c=1/9$. On the other hand, in the spinful model  ($N_s=2$) the transition into the CDW$_3$ phase occurs at a critical value 
 $(\tilde{v}_2)_c = 1/12$, pre-empting the transition into the QSH phase. These findings are in qualitative agreement with previous mean field 
 studies on the lattice \cite{grushinetalprb2013,kuritaetalprb2016,bijelicetalprb2018}.

{\it Semimetallic charge order.}$-$ In the absence of NN repulsion, the favored charge-ordered state with $\rho> 0$ and $\Delta=0$ describes a hidden smectic order with gapless excitations. 
This broken-symmetry state remains \emph{semimetallic}, with one pair of bands opening a gap and another pair remaining gapless, as shown in 
Fig.~\ref{semi-Dirac fermions+folded BZ}(b).
The effective Hamiltonian matrix of the two gapless bands in the CDW$_3$ phase is 
\begin{equation}
\hat{\mathcal{H}}(\mathbf{p}) =  v_F p_x\tau^x  +  v_F^2 p_y^2/(4 V_2 \rho) \tau^z, 
\end{equation}
with energy spectrum $|\varepsilon_\pm(\mathbf{p})| =  v_F \sqrt{p_x^2+ v_F^2 p_y^4/(4 V_2\rho)^2 }$. 
The quasiparticles are  {\it semi-Dirac} fermions, which disperse linearly along the $x$ direction and have a parabolic touching along the $y$ axis. Those touching points sit at the high symmetry $\Gamma$ points of the folded Brillouin zone (see Fig.~\ref{semi-Dirac fermions+folded BZ}).

{\it Fluctuations effects.}$-$
Fluctuation corrections to the topological Mott order are best captured by decoupling the interaction in the bond-order channel by means of a 
Hubbard-Stratonovich transformation. The resultant action $S\sim\int_{\tau,\v{r}}\bar{\psi}(\hat{G}^{-1}_0+iV_2s^\mu\hat{\chi}^\mu)\psi+V_2\hat{\chi}^\mu\hat{\chi}^\mu$
is quadratic in the fermionic Grassmann fields $\bar{\psi},\psi$ at the expense of introducing imaginary collective bosonic fields $i\hat{\chi}^\mu_\sigma$.
Both vary in position $\v{r}$ and imaginary time $\tau$.

We formulate a self consistent expansion around the broken-symmetry states. This is equivalent to working with the 
renormalized propagator $\hat{G}^{-1}=\hat{G}_0^{-1}+\hat{\Sigma}$, where  $\hat{G}^{-1}_0=\hat{\partial}_\tau-\hat{H}_t$ is the bare fermionic propagator and
 $\Sigma_{s}=\frac{3\sqrt{3}}{2}V_2 \chi_s\sigma^z \nu^z$ the self energy due to the zero frequency fields $\chi_s= \chi^0$ or $s\chi^z$ for the QAH and QSH phases respectively, with $s=\pm$ indexing the spin. Inclusion of the finite frequency fluctuation fields 
$\tilde{\chi}^\mu_\sigma$ amounts to the addition of a Yukawa coupling to the low-energy effective action, $\mathcal{S}=\mathcal{S}_\psi+\mathcal{S}_{\tilde{\chi}}
+\mathcal{S}_{\psi\tilde{\chi}}$, with
\begin{align}
\mathcal{S}_{\psi\tilde{\chi}}&=\frac{3\sqrt{3}}{2}V_2\sum_{\nu\sigma}\int\dbar^3\vec{p}_1\dbar^3\vec{p}_2\,
\bar{\psi}_{\vec{p}_1\sigma}^\nu\tilde{\chi}_{\vec{p}_1-\vec{p_2}\sigma}^{\mu}
s^\mu\psi_{\vec{p}_2\sigma}^\nu.
\end{align}
Here $\vec{p}=(v_F\v{p},\omega)$, $\dbar^3\vec{p}=\dbar^2\v{p}\,d\omega/(2\pi v_F^2)$, and $\nu=\pm$ indexes the valleys. 

\begin{figure}[t]
\includegraphics[width=0.95\columnwidth]{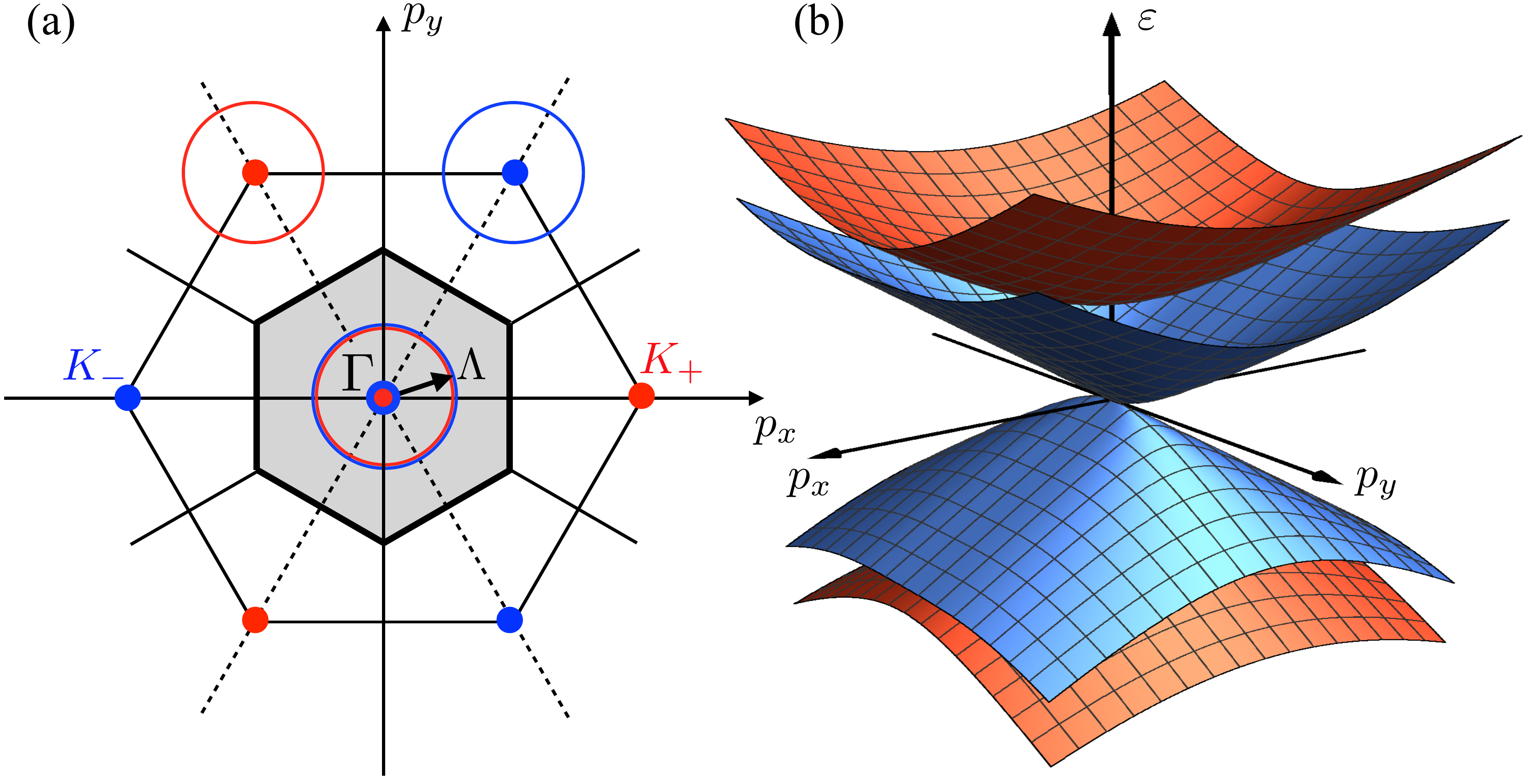} 
\caption{ \label{semi-Dirac fermions+folded BZ}
(color online) (a) Gray area: folded Brillouin zone in the CDW$_3$ state. Valleys in the normal state (red and blue dots) fold into the center of the zone. (b) Low energy bands of the CDW$_3$ state around the $\Gamma$ point. At half filling, the system is gapless, with semi-Dirac quasiparticles.}
\end{figure}

 Integration over the fermionic fields to quadratic order in $\tilde{\chi}$ yields the fluctuation 
action $\tilde{\mathcal{S}} =\mathcal{S}_{\tilde{\chi}}-\frac{1}{2}\avg{\mathcal{S}^2_{\psi\tilde{\chi}}}
=\int_{\vec{q}}\tilde{\chi}^{\mu}_{-\vec{q}\sigma}A^{s^\prime}_{\vec{q}\sigma\sigma^\prime}\tilde{\chi}^{\mu}_{\vec{q}\sigma^\prime}$, 
which decouples into the longitudinal $\tilde{\chi}^{0}$, $\tilde{\chi}^{z}$ ($s^\prime=s$) and transverse $\tilde{\chi}^{x}$, $\tilde{\chi}^{y}$ ($s^\prime=-s$) sectors.
The matrix elements
\begin{align}
A_{\sigma\sigma^\prime}^{s^\prime}(\vec{q})=\frac32 V_2\left(\delta_{\sigma\sigma^\prime}+\frac94 \gamma{}V_2\sum_s\Pi^{ss^\prime}_{\sigma\sigma^\prime}(\vec{q})\right)\label{tmiquadraticfluctuationcorrection}
\end{align}
depend on the fermionic polarization bubbles $\Pi^{ss^\prime}_{\sigma\sigma^\prime}(\vec{q})=\sum_\nu\int\dbar^3\vec{p}\,
G^{\nu{}s}_{\sigma\sigma^\prime}(\vec{p}+\vec{q})G^{\nu{}s^\prime}_{\sigma^\prime\sigma}(\vec{p})$ for the broken symmetry states. In matrix form, $\Pi^{s s^\prime}\!(\vec{q}) = \Pi^{s s^\prime}_\mu \!(\vec{q})\sigma^\mu$,  where 
\begin{eqnarray}
  \Pi^{s s^\prime}_{0}\!(\vec{q}) & \approx & \frac{\lambda}{q}
      \left( q^2+\theta^2+ 4M^2\frac{q^2-\theta^2}{q^2}+8M_sM_{s^\prime}\right), \quad\\
 \Pi^{s s^\prime}_x(\vec{q})\! & \approx & -\frac{\lambda}{q}\left(
    2 q^2 +v_F^2\mathbf{q}^2 + 4M^2\frac{    v_F^2\mathbf{q}^2-2q^2}{q^2}
    \right). \quad
 \end{eqnarray} 
up to second order in $M_s = 3\sqrt{3}/2V_2 \chi_s$, with $\lambda=\pi^2/(8v_F^2A)$, $\vec{q}\cdot\vec{q}=q^2$ and $\Pi_y^{s s^\prime}(\vec{q})=\Pi_z^{s s^\prime}(\vec{q})=0$ \cite{note}.

The constant  $\gamma$ in Eq.~(\ref{tmiquadraticfluctuationcorrection}) is a phenomenological parameter that has been included to account for renormalization of the 
vertex $V_2\tilde{\chi}\bar{\psi}\psi$ from:  (i) coarse-graining the lattice in a Wilsonian sense; (ii) higher order $\tilde{\chi}$ terms;
(iii) the Fermi velocity renormalization as $\Pi\propto{}1/v_F$. Both the theoretical and experimental evidence for graphene
\cite{gonzalezetalnpb1994,eliasetalnature2011,kotovetalrmp2012}  suggests $\gamma <1$.  In addition, $\gamma$ has the added benefit of smoothly 
interpolating between mean field ($\gamma=0$) and the bare coupling with fluctuations $(\gamma=1)$.

The Gaussian integrals over the fluctuation fields lead to the free-energy corrections $\delta{}f_{s^\prime}=\text{Tr}\ln{}A^{s^\prime}$, 
from which we obtain the fluctuation contributions to the quadratic coefficients of the Landau expansion,
\begin{align}
  \delta\alpha^\mu_{s^\prime}=\frac{1}{2}\int\dbar^3\vec{q}\;\text{Tr}\;
  \frac{\gamma{}V_2\sum_s\partial_{\chi^\mu}^2\Pi^{ss^\prime}\!(\vec{q})}
  {\sigma^0+\gamma{}V_2\sum_s\Pi^{ss^\prime}\!(\vec{q})}\bigg|_{\chi^\mu=0}.
\end{align}
Remarkably,  it is possible to evaluate the expressions analytically. For the QSH order parameter we obtain   
\begin{align}  
    \delta{}\alpha^z_L&=\frac{24V_2}{\gamma\pi^2}\left(\text{arccot}^2 \Omega
                  -\frac12\ln\frac{\Omega^2+3}{\Omega^2+1}\right),\\
  \delta{}\alpha^z_{T}&=-\frac{54 v_2}{\pi}V_2\left(1-\Omega\,
                        \text{arccot}\Omega\right),
\end{align}
for the contributions from longitudinal and transverse fluctuations, where $\Omega = \sqrt{8/(9\pi\gamma{}v_2)-1}$.
For the QAH order we obtain  
$\delta\alpha_L^0=\delta\alpha_T^0=\delta\alpha_L^z$.   The calculation breaks down for 
$v_2\ge{}8/9\pi\gamma$. 

In the case of the CDW$_3$ state, the interaction is decomposed in the charge channel by introducing six auxiliary fields 
$\hat{\rho}_i$ ($i=1,\ldots,6$), one for each site in the extended unit cell,
\begin{align}
&\sum_{\avg{\avg{i,j}}}\hat{n}_i\hat{n}_j =\int_{\vec{q}}\sum_{i,j}\hat{n}_{-\vec{q}i}U^{ij}_{\v{q}}\hat{n}_{\vec{q}j} \rightarrow \notag\\
&\int_{\vec{q}}\sum_{i,j}\rho_{-\vec{q}i}(U^{-1}_{\v{q}})^{ij}\rho_{\vec{q}j}+ 2\int_{\vec{k}\vec{q}} \sum_{i}\bar{\psi}_{\vec{k}+\vec{q}i}s^0\psi_{\vec{k}i}\rho_{\vec{q}i}.
\end{align}
After projecting into the low energy fermionic subspace, the calculation proceeds as before. First we expand around the mean field solutions, 
$\rho_i \to \langle \hat{n}_i \rangle +\tilde{\rho}_i$, which 
dress the fermion propagator. We then integrate over the fermionic fields to obtain the Gaussian action 
$\tilde{\mathcal{S}}=\int_{\vec{q}}\sum_{ij}\tilde{\rho}_{-\vec{q}i}\tilde{A}_{\vec{q}}^{ij}\tilde{\rho}_{\vec{q}j}$ for the finite-frequency charge fluctuations, 
where $\tilde{A}^{ij}_{\vec{q}} = (U^{-1}_{\vec{q}})^{ij} +2\gamma V_2 N_s \tilde{\Pi}_{\vec{q}}^{ij}$, with $\tilde{\Pi}_{\vec{q}}^{ij}$ the charge polarization tensor in the CDW$_3$ phase. Integration over the fluctuation fields gives the leading 
free energy corrections $\delta\tilde{f}=\mathrm{Tr} \ln\tilde{A}$ in terms of the order parameters $\rho$ and $\Delta$. In general, the coefficients of the 
expansion can be evaluated numerically. Approximating $U_{\v{q}}\approx U_{\v{q}=\v{0}}$, the quadratic coefficient 
$\delta\tilde{\alpha}(x)$ ($x=\Delta/\rho$) can be obtained in
analytic form \cite{supplementalmaterial}. The resulting 
phase boundaries are almost identical to the ones obtained from numerical integration. 
We find that fluctuations do not change the nature of the charge order: the CDW$_3$ state remains \emph{metallic} with $\Delta=0$ ($x=0$).

\begin{figure}[t]
  \includegraphics[width=0.9\columnwidth]{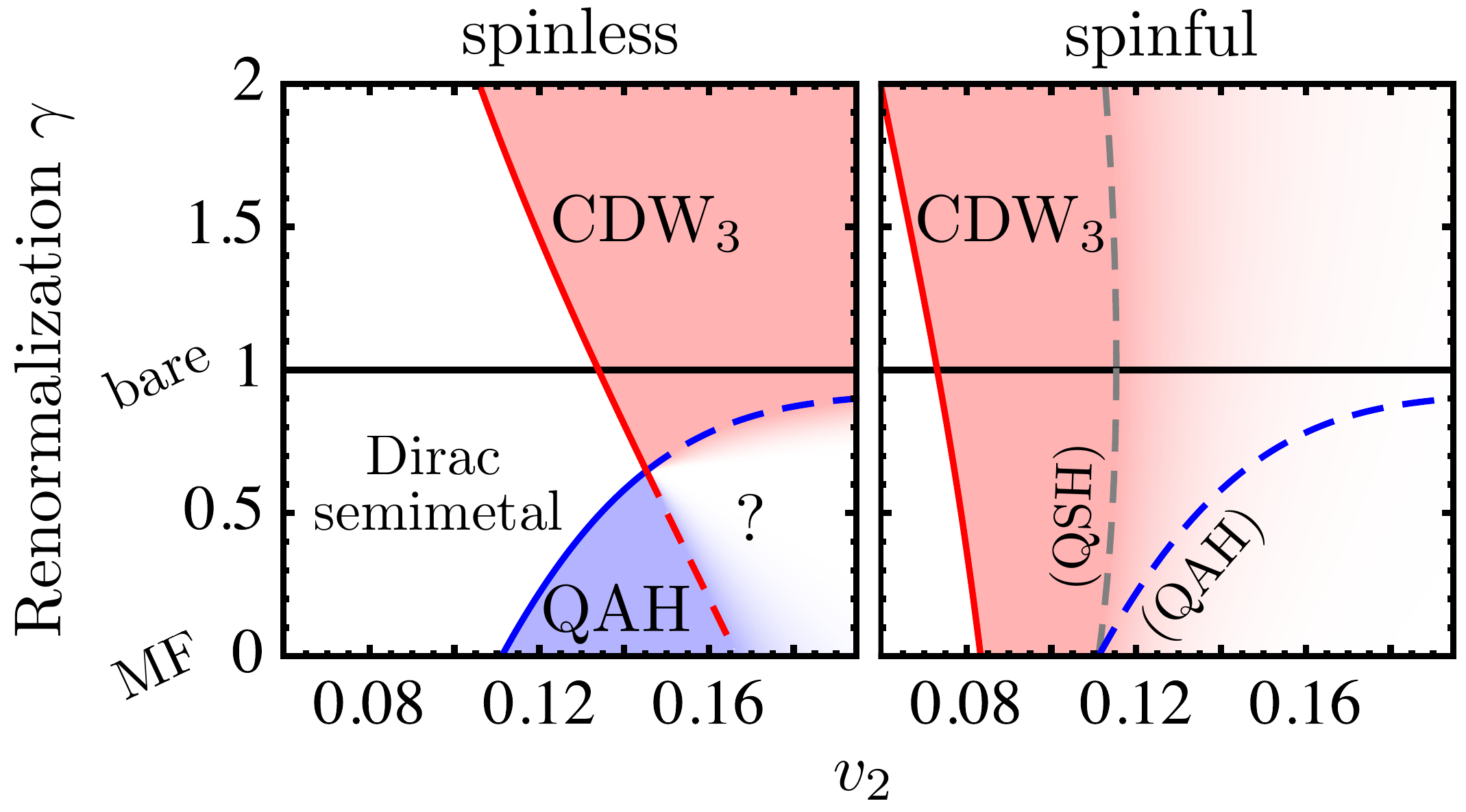} 
\caption{\label{phasediagramfigure}
Lines of critical-instability along the $v_2=\frac{\pi\Lambda}{v_FA}V_2$  axis in the presence of fluctuations, renormalized by the 
phenomenological parameter $\gamma$. The mean field instabilities are at $\gamma=0$, 
the cut $\gamma=1$ indicates the phase behavior without vertex renormalization. While the  critical interaction strengths depend on the momentum cut-off $\Lambda$,
the order of instabilities does not.  In the regime where the NN interactions are zero ($V_1=0$), the CDW$_3$ 
phases are gapless ($\Delta=0$). }
\end{figure}

{\it Phase diagram and discussion.}$-$ Our main results are summarized in Fig.~\ref{phasediagramfigure}. For the spinless model the leading instability at mean field 
($\gamma=0$) is to the topological QAH Mott insulator. Fluctuations favor CDW$_3$ order  over the QAH state and are strong 
enough to cause a continuous phase transition from the Dirac semimetal to the CDW$_3$ phase for $\gamma\gtrsim 0.62$.
This is precisely the nature of the transitions found within numerical approaches \cite{daghoferhohenadlerprb2014,motruketalprb2015,capponilauchliprb2015,schereretalprb2015}. 
Similar fluctuation-driven changes of the ground state have been recently discussed in terms of a fermionic quantum order-by-disorder mechanism 
\cite{Kruger+12,Kruger+14,AbdulJabbar+2015,Green+18}. In the spinful model the transverse fluctuations in the QSH phase stabilize the order, lifting the mean-field degeneracy 
of the QSH and QAH phases,  $\delta\alpha_T^z<0<\delta\alpha_L^z=\delta\alpha_{T/L}^0$. The transverse fluctuations are not strong 
enough however to suppress the CDW$_3$ phase, which is the leading instability at mean-field. 

The transition to the gapless CDW$_3$ state ($\rho>0$, $\Delta=0$) is highly unconventional since the ground state remains metallic with semi-Dirac quasiparticles.
It does not belong to the class of putative Dirac semimetal-to-insulator transitions. Instead, the fermion residue remains finite across the transition.
This hidden charge order eluded previous numerical studies \cite{garciamartinezetalprb2013,daghoferhohenadlerprb2014,djuricetalprb2014,capponilauchliprb2015,motruketalprb2015, 
schereretalprb2015,volpezetalprb2016,delapenaetalprb2017, kuritaetalprb2016,bijelicetalprb2018} that identified phase transitions through the 
opening of a Mott gap. The onset of semi-Dirac behaviour may be resolved in large-scale DMRG simulations on infinite cylinders, which are now capable of extracting 
the momentum-dependent excitation spectra of Dirac materials \cite{He+17}. Finally, with the recent advent of  ``designer 
Hamiltonian'' methods \cite{satoetalprl2017,heetalprb2018} in quantum Monte Carlo it seems possible to engineer the unconventional
self-energy terms of the CDW$_3$ state.

By modifying the renormalization-group studies of GNY models \cite{zinnjustinnpb1991,rosaetalprl2001,herbutetalprb2009,janssenherbutprb2014}, it will be possible to 
unravel the nature of the quantum critical point and its stability against other couplings. 
 As we demonstrated, the hidden CDW$_3$ order 
is stable against Gaussian fluctuations. We believe that this stability holds under the RG since the NNN coupling $V_2$ does not generate interactions 
between the sublattices that would lift the degeneracy underlying the quadratic touching.  

A small NN repulsion $V_1$ leads to the opening of a Mott gap. Closer inspection shows that the semi-Dirac mode splits into two massive 
Dirac cones along the quadratic touching direction.  
 While in this case the  transition is likely to belong 
to the chiral Ising GNY universality class, we expect to see a characteristic 
crossover in the critical fluctuations due the proximity to the unusual critical point at $V_1=0$. It has been suggested  \cite{bijelicetalprb2018} that the regime of 
dominant $V_2$ could become experimentally accessible by using silicon adatoms or cold atoms in double-layers of triangular optical lattices. 
 
In materials with a quadratic band-touching, such as bilayer graphene \cite{McCann+06}, interactions are marginally relevant \cite{Sun+09}.  Linear terms in the dispersion are generated under the RG, pushing the critical interaction 
strength back to a finite value and leading to GNY universality \cite{Pujari+16}. In our case, the bare electron dispersion is already linear. Only because of the matrix 
structure of the Yukawa coupling for $V_1=0$, the symmetry breaking does not lead to the opening of a gap but instead to a quadratic touching along the CDW$_3$ order.

{\it Acknowledgements.}$-$ We thank Andrew Green,  Andrew James and Fernando de Juan for useful discussions. B.~U. acknowledges NSF CAREER grant 
No. DMR-1352604 for partial support. F.~K. acknowledges financial support from EPSRC under Grant EP/P013449/1.

\end{document}